\documentclass[manuscript]{aastex}
\usepackage{epsfig}
\usepackage{amsmath}
\begin{document}

\title{Singlet Excited States of Anions with Higher Main Group Elements}

\date{\today}
\author{Ryan C. Fortenberry}
\email{rfortenberry@georgiasouthern.edu}
\affil{Department of Chemistry, Georgia Southern University, Statesboro, GA 30460, U.S.A.}

\begin{abstract}

Previous studies have shown that dipole-bound excited states exist for certain
small anions.  However, valence excited states have been reported for some
closed-shell anions, but those with singlet valence excited states have, thus
far, contained a single silicon atom.  This work utilizes high-level coupled
cluster theory previously shown to reproduce excited state energies to better
than 0.1 eV compared with experiment in order to examine the electronic excited
state properties of anions containing silicon and other higher main group atoms
as well as their first row analogues.  Of the fourteen anions involved in this
study, nine possess bound excited states of some kind: CH$_2$SN$^-$,
C$_3$H$^-$, CCSiH$^-$, CCSH$^-$, CCNH$_2^-$, CCPH$_2^-$, BH$_3$PH$_2^-$,
AlH$_3$NH$_2^-$, and AlH$_3$PH$_2^-$.  Two possess clear valence states:
CCSiH$^-$ and its first row analogue C$_3$H$^-$.  Substantial mixing appears to
be present in the valence and dipole-bound characters for the first excited
state wavefunctions of many of the systems reporting excited states, but the
mixing is most pronounced with the ammonia borane-like AlH$_3$NH$_2^-$, and
AlH$_3$PH$_2^-$ anions.  Inclusion of second row atoms in anions whose
corresponding radical is strongly dipolar increases the likelihood for the
existence of excited states of any kind, but among the systems considered to
date with this methodology, only the nature of group 14 atoms in small,
closed-shell anions has yet been shown to allow valence singlet excited states.

\end{abstract}

\maketitle


\section{Introduction}

In the search for carriers of unexplained interstellar absorption spectra, the
electronic properties of rarely studied systems offer a viable new avenue of
research to explore.  The set of as-of-yet unexplained visible to near-infrared
peaks known as the diffuse interstellar bands (DIBs), for example, has remained
a mystery for nearly a century \citep{Sarre06}. However, it has been suggested
that anions may be responsible for some of the DIB features \citep{Sarre00,
Cordiner07}.  Recent work has shown a relative richness of electronic absorption
features below the electronic binding energy (eBE) \citep{Simons08, Simons11} for
some anions that exhibit singlet dipole-bound excited states and even some
species that appear to possess additional valence
excitations \citep{Fortenberry11dbs, Fortenberry112dbs} For any dipole-bound
anion, the corresponding radical must possess large a dipole
moment \citep{Fermi47, Turner66, Coulson67, Crawford67, Jordan76, Turner77,
Crawford77, Gutowski96, Wang02, Jordan03} in order for the electron to remain
bound within the system.  In an ideal situation the dipole moment must be at
least greater than 1.625 D \citep{Fermi47, Turner66, Coulson67} and probably
closer to 2.5 D \citep{Cordiner07, Gutsev95a, Ard09} typically in a highly
diffuse orbital \citep{Mead84}.  However, valence ground and excited states may
exist below the single dipole-bound state which functions as the upper
excitation energy limit since the dipole-bound state excitation energy must be
nearly coincident with the eBE \citep{Simons08}.


``Bound excited valence states in molecular anions are rare" \citep{Brinkman93}.
However, some anions are known to exhibit valence electronic excitations, but
most of these are open-shell anions where excitations do not have to overcome
the additional energy barrier of sorts involved in splitting an electron pair
while still remaining below the eBE.  Negatively charged small fragments or
atoms \citep{Brinkman93} and, especially, carbon chains \citep{Maier98,
Jochnowitz08, Zhao09} have been documented to possess bound valence excited
states.  Additionally, some anions, including many with second row atoms, are
known to possess triplet excited states above the ground state \citep{Meloni04,
Pino04, Sheehan08, Inostroza10} where the energy splitting between the ground
state (which may either be singlet or triplet) and the triplet excited state is
relatively small.  Much larger anions are also known to possess excited
states, \citep{Brinkman93, Skurski00TCNQ, Sobczyk03} and such behavior is
relevant to the study of many biological species \citep{Chen01}. However, a new
set of small, closed-shell singlet anions have been studied previously for
their electronic properties \citep{Fortenberry11dbs, Fortenberry112dbs}.  They
represent an emerging class of anions that are not: 1) large, long, or exotic,
2) atomic, elemental, or diatomic, or 3) radical, diradical, or open-shell.
They are closed-shell anions of three to seven atoms (with no more than four
heavy atoms) that are composed of common substituent groups.

Silicon is present in each of the small singlet systems where previous study
indicates the existence of valence excited states \citep{Fortenberry11dbs,
Fortenberry112dbs}. Silicon and other higher row main group elements are
well-known to have very different bonding environments than their more common
C, O, and N (first row) analogues \citep{Kutzelnigg84, Driess96, Power99,
Brown00, Owens06, Woon09SFn, Woon10PFn, Woon11}. Silicon provides many
unexpected differences in the molecular structures as compared to carbon
molecules of the analogous chemical formula including the butterfly conformer
of Si$_2$H$_2$ and cyclic Si$_3$ \citep{Li95, Chesnut02, Reilly12}. Steric
effects from the presence of the larger $n=3$ orbitals in the higher row atoms
inhibit standard valence bond $sp^3$/$sp^2$ hybridization or electron pair
recoupling \citep{Woon11} making molecules containing these atoms ``reluctant to
hybridize" \citep{Kutzelnigg84, Driess96, Owens06}. The lack of hybridization is
most extreme in second row atoms, most notably for silicon, which forms strong
$\sigma$ bonds due to an increase in its inclusion of the $s$-type atomic
orbitals \citep{Power99, Brown00, Owens06}.  Hence, silicon is even less likely
to create bonds of significant hybridized character than its rowmates
phosphorus and sulfur.  The increase in $s$ character is most straightforwardly
observed in $^1A_1$ SiH$_2$ where the optimal bond angle is
92.1$^{\circ}$, \citep{Dubois68, Driess96} much closer to 90$^{\circ}$ than
102$^{\circ}$, which the bond angle for singlet carbene \citep{Bourissou00}. Even
so, the carbene carbon here exibits similar effects since the bond angles in
both $^1A_1$ CH$_2$ and SiH$_2$ are much less than the 120$^{\circ}$ bond angle
expected for standard $sp^2$ hybridized atoms \citep{Woon11}.

The previous work examining the singlet excited states of the aforementioned
small, closed-shell anions \citep{Fortenberry11dbs} employed coupled cluster
methods to reproduce the experimental eBEs and the lone dipole-bound state
excitation energies of CH$_2$CN$^-$ \citep{Sarre00, Cordiner07, Lykke87} and
CH$_2$CHO$^-$ \citep{Mead84, Mullin92, Mullin93} to within 0.06 eV or better.
The same accurate methodology was extended to provide likely dipole-bound and
valence excited states of other anions for which experiments have not yet been
performed.  The present study employs the same methodology and is an
examination of the excited state properties of similar molecular anions
containing the second row elements aluminum, silicon, phosphorus, and sulfur
and how they relate to their first row analogues.  Some systems analyzed have
been constructed to produce structures related to those where the previous
computational studies indicate the presence of dipole-bound and valence excited
states for silicon-containing species \citep{Fortenberry11dbs,
Fortenberry112dbs}.  Other anions of interest to this work are unique attempts
to find if the larger atoms play a role in the stabilization of valence
excitations below the eBE.  The search for dipole-bound states provides
motivation for analysis of these types of systems, but the search for even
rarer valence states offers a more fascinating result.  It is currently unknown
how anions may affect attribution of the DIBs, but the probable existence of 
excited states for previously unexamined molecules may hold some new insight
into this near-century-old problem.


\section{Computational Details}

Coupled cluster theory is one of the most accurate quantum chemical methods
developed to date, \citep{Helgaker04, Lee95Accu, Shavitt09} and our
past \citep{Fortenberry11dbs, Fortenberry112dbs} and current computations of
excited states of anions exclusively utilize this method extensively.  The
geometries are optimized at the coupled cluster singles, doubles, and
perturbative triples level [CCSD(T)] \citep{Ragh89} in conjunction with Dunning's
singly augmented correlation consistent triple-zeta basis set, aug-cc-pVTZ
\citep{Dunning89, cc-pVXZ} and aug-cc-pV(T+d)Z for the higher row
atoms \citep{Dunning01}.  Spin-restricted (RHF) \citep{ScheinerRHF87, Lee91}
reference wavefunctions are used in computations involving the closed-shell
anions while spin-unrestricted (UHF) \citep{Gauss-UHF91, Watts-UHF92}
wavefunctions are chosen for the open-shell radical computations.  Dipole
moments are computed at the UHF-CCSD/aug-cc-pVTZ level of theory from the
UHF-CCSD(T)/aug-cc-pVTZ optimized geometry of the radical.

The computation of vertically excited states from the RHF-CCSD(T)/aug-cc-pVTZ
{\footnote{The aug-, n-aug-, and cc-pV(X+d)Z basis sets for the higher row
atoms are simply referred to as the aug-, n-aug-, and cc-pVXZ basis sets beyond
the Computational Details section for ease of discussion.}} reference geometry
makes use of equation of motion coupled cluster (EOM-CC) theory
\citep{Stanton93EOM, Monkhorst77, Mukherjee79} with increasingly diffuse basis
sets constructed in an even-tempered fashion.  Inclusion of this series of
basis sets provides evidence as to the classification of whether a state is
valence or dipole-bound.  A precipitous decrease in excitation energy from a
standard cc-pVXZ (where X=D or T) basis set \citep{Dunning89, cc-pVXZ,
Dunning01} to t-aug-cc-pVXZ indicates the presence of a dipole-bound state.
However, such is only the case if the excitation energy is below or within a
computational limit (usually 0.1 eV) of the vertical eBE
\citep{Fortenberry112dbs} computed with the same level of theory and basis set
for the reference geometry.  These vertical eBEs are computed with the equation
of motion coupled cluster theory for ionization potentials (EOMIP)
\citep{Stanton94:EOMIP} approach where an electron is removed from the HOMO of
the anion.  A valence state must also be less energetic than the vertical eBE,
but its transition energy is typically much less than this limit and certainly
below that of the dipole-bound state excitation energy \citep{Fortenberry11dbs,
Simons08}. Additionally, the decrease in vertical excitation energy with more
diffuse basis sets is much less striking for a valence state where nearly all
of the necessary orbital extent is recovered with the use of the aug-cc-pVXZ
basis \citep{Skurski00}.  Convergence is thus clearly achieved for computations
involving the other, more highly diffuse bases.

Adiabatic computations of the eBEs and excitation energies further provide
insight into the potential existence of either type of excited state since they
allow for a more direct comparison to experiment \citep{Fortenberry11dbs,
Fortenberry112dbs}.  Adiabatic eBEs represent the energy difference between
CCSD(T)/aug-cc-pVTZ optimized geometries of the neutral and the
anion \citep{Simons11}.  Similarly, the adiabatic excitation energies are the
energy difference between the CCSD/d-aug-cc-pVDZ optimized ground state and the
EOM-CCSD/d-aug-cc-pVDZ optimized excited state.  Also, the optimized geometries of
the ground state radicals and dipole-bound excited states should be very
similar, which gives a further means by which excited states can be classified.

Core orbitals are frozen in all computations: $1s^2$ for B, C, N, and O with
$1s^2 2s^2 2p^6$ frozen for Al, Si, P, and S.  All geometry optimizations are
minima according to harmonic vibrational frequency analyses unless otherwise
noted and discussed.  The EOMIP-CC computations make use of the
{\footnotesize{CFOUR}} \citep{CFOUR} quantum chemical program package, while all
other computational results were obtained using the
{\footnotesize{PSI}}{\small{3}} suite of computational chemistry
programs \citep{Psi3}.

\section{Results and Discussion}

The presence of valence excited states in silicon-containing anions from our
previous studies \citep{Fortenberry11dbs, Fortenberry112dbs} has led us to to
explore if silicon's properties are unique in the retention of anionic excited
states, especially valence states, or if the general properties of higher row
atoms allow for behavior similar to that previously computed.  The systems
chosen for this study all have closed-shell ground states that are valence in
nature.  In order to meet this requirement, the first set of systems considered
have a methylene group on one end of the structure and a single ``dipole moment
inducing" atom (labeled as M) on the other where the two are bridged by our
atom of interest, X.  These anions have the form CH$_2$XM$^-$ since
CH$_2$SiN$^-$ appears to possess both valence and dipole-bound excited states.
In the CH$_2$XM$^-$ systems, X=N, P, O, and S which compare with X=C and Si
from \citep{Fortenberry11dbs}.  M equals N when X is either O or S, but
M=O where X=N and P.  The choice of M is limited to first row atoms of groups
15 or 16 due to electronegativity effects, which influence dipole moment
strength, and the lack of dipole-bound states previously computed for fluorine
containing anions \citep{Fortenberry11dbs}.

The second set of systems studied are carbide groups paired with XH$_n$ groups.
Here the X atoms studied are C, N, O, Si, P, and S.  The $n$ values are 2 when
X=N and P, and $n=1$ for all other systems.  This construction gives six
different anions including C$_3$H$^-$ and CCSiH$^-$.  These are chosen since
the carbide containing CCSiN$^-$ has been shown to possess two valence excited
states, \citep{Fortenberry112dbs} and we are exploring how the carbide moiety can
affect atoms in both rows without the inclusion of further influence from the
additional N atom present in CCSiN$^-$.

The third class of systems in this study is the $^1A'$ XH$_3$YH$_2^-$ systems
where X=B or Al and Y=N or P.  Related systems, with BH$_3$NH$_3$ being the
most notable, have been the subject of much computational study, especially
with regards to the nature of the X$-$Y bond length, stretching frequency, and
dissociation \citep{Thorne83, Binkley83, Marsh92, Leboeuf95, Reinemann11,
Sams12}. Presently, this system allows us to probe anions with group 13 elements
and further explore group 15 elements.  The group 15 atoms of the
XH$_3$YH$_2^-$ systems are chosen to be the Y atoms in the YH$_2$ groups.  This
arrangement creates dipole moments in the radicals large enough to allow
potentially for dipole-bound states.  Additionally, some anionic and
corresponding radical systems with the group 13 atoms in YH$_2$ are not stable
structures.

\subsection{Excited States of CH$_2$XM$^-$ Anions}


Table \ref{ebe} and Table \ref{basis} provide a summary of the data for each of
the systems of interest for this work.  The first system similar to
CH$_2$CN$^-$ and CH$_2$SiN$^-$ listed in Table \ref{ebe} is CH$_2$ON$^-$, which
is also shown in Figure \ref{ebeFig}a \& Figure \ref{AdFig}a.  CH$_2$ON$^-$ is
simply a replacement of the cyano carbon with an oxygen to create an anion
similar to CH$_2$CN$^-$.  Both the neutral radical and the anionic forms of
CH$_2$ON are asymmetric molecules as has been shown previously for the
radical \citep{Shapley98}.  The adiabatic eBE for CH$_2$ON$^-$ cannot actually be
determined here since the $C_1$ computations are not well-behaved in the SCF
step.  However, it can clearly be determined that the eBE will be lower than
0.46 eV, which is the adiabatic eBE from consideration of the planar CH$_2$ON
radical and the optimized asymmetric anion.  This upper bound negates the
existence of any stable excited states for CH$_2$ON$^-$ and is much lower in
energy than the vertical eBE. Even though two vertically excited states are
computed to lie below the vertical eBE, only one of these dipole-bound states
could exist if the adiabatic eBE was much higher \citep{Fermi47, Simons08}.



Isomerization of CH$_2$ON$^-$ to CH$_2$NO$^-$ results in a species that is more
energetically favorable.  The computations report that CH$_2$NO$^-$ is 75.0
kcal/mol more stable than the CH$_2$ON$^-$ isomer.  The corresponding radicals
are similarly separated by around 50 kcal/mol, which is in agreement with
previous computations at various levels of theory \citep{Shapley98, Shapley99}.
These two anions have comparable dipole moments of $\sim 2.3$ D as shown in
Table \ref{ebe}.  CH$_2$NO$^-$ and its excited state properties have been
previously examined by \cite{Fortenberry11dbs}, but they are relevant for
this discussion.  The 1.55 eV adiabatic excitation energy to the lowest excited
state (2 $^1A'$) is more than 0.1 eV higher than the 1.42 eV adiabatic eBE,
which indicates that this state does not exist.  However, vertical excitation
energies from the RHF-CCSD(T)/aug-cc-pVTZ optimized geometry reported in Table
\ref{basis} show a clear pattern of energy decrease with the inclusion of more
diffuse functions indicative of a highly diffuse dipole-bound state.
Interestingly, the vertical CCSD/t-aug-cc-pVTZ eBE computed via EOMIP is much
higher than the vertical EOM-CCSD/t-aug-cc-pVTZ 2 $^1A'$ state energy.
However, the adiabatic results better model physical behavior and this leads to
the conclusion that, again, CH$_2$NO$^-$ probably does not possess any bound
singlet excited states.

Phosphorus substitution for the nitrogen atom in CH$_2$NO$^-$ to give
CH$_2$PO$^-$ and sulfur substitution of the oxygen atom in CH$_2$SN$^-$ both
appear to give dipole-bound states.  CH$_2$PO radical's dipole moment of 2.477
D is well above Fermi and Teller's \citep{Fermi47} theoretical 1.625 D necessary
dipole moment for a dipole-bound state in the corresponding anion and right at
Gutsev and Adamowicz's \citep{Gutsev95a} more practical 2.5 D limit.  The 2.477 D
dipole moment for CH$_2$PO is higher than either of the previous first row
analogues, indicating that presence of the second row atom has some effect on
the system.

The adiabatic transition energy to the first excited state of CH$_2$PO$^-$, 1
$^1A''$, is computed to be 2.89 eV, while the eBE is 2.82 eV.  This is within
the 0.1 eV limit of computational accuracy.  Vertically, the eBE is 2.79 eV,
and the vertical excited state energy for the t-aug-cc-pVDZ basis set is the
same.  Coupling these factors with a steep decline in the vertical excitation
energy for increasingly more diffuse basis sets shown in Table \ref{basis}
gives classic behavior of a dipole-bound excited state.  However, the
excitation into a highly diffuse $s$-type (totally symmetric) orbital
necessitates that the term for the dipole-bound excited state must be directly
related to the term of the radical, $^2A'$ in this case.  There is a second
excited state of CH$_2$PO$^-$, 2 $^1A'$, that has an adiabatic excitation
energy (3.08 eV) much higher than 2.82 adiabatic eBE.  Its t-aug-cc-pVDZ
vertical excitation energy of 2.81 eV is only 0.02 eV higher than the vertical
eBE.  Another key hallmark of a dipole bound state is that due to the
diffuseness of the orbital in which the bound electron is held, the excited
anion will take on not only the similar term but also the geometry of the
radical whose dipole moment is binding the electron.  Figure \ref{AdFig}d shows
that the 2 $^1A'$ CH$_2$PO$^-$ geometry is very much like the Figure
\ref{ebeFig}d CH$_2$PO geometry.  Additionally, only one dipole bound state can
exist for any given anion \citep{Fermi47, Simons08}.  Hence, the 2 $^1A'$ state
is actually the dipole-bound state, but the adiabatic results for this
dipole-bound excited state do not suggest that it will be stable since it is
more than 0.1 eV higher than the eBE.


The 1 $^1A''$ state of CH$_2$SN$^-$ behaves somewhat differently than the
analogous 3 $^1A$ state of CH$_2$ON$^-$.  There is relatively little change in
the vertical excitation energy (0.72 eV) from cc-pVDZ to t-aug-cc-pVDZ for the
CH$_2$SN$^-$ 1 $^1A''$ state, and the vertical eBE at 2.24 eV is higher than
the t-aug-cc-pVDZ excitation energy at 2.04 eV.  However, the change from d- to
t-aug-cc-pVDZ excitation energy is fairly substantial at 0.39 eV.  This is over
half of the total decrease in excitation energy across the range of diffuse
basis sets indicating that there is a mixture of valence and dipole-bound
character.  Even so, there is not enough pure valence character in the
excitation in order for this state to be predominantly valence in nature.
Similar vertical CC3 excited state computations also do not give any further
evidence to show that the 1 $^1A''$ state is a valence state.  Additionally,
the adiabatic excitation energy for this state is more than 0.1 eV higher than
the adiabatic eBE.  As a result, we must conclude that the 1 $^1A''$ state of
CH$_2$SN$^-$ is not accessible.  Differently, the lower $2\ ^1A'$ state is the
dipole-bound state since the adiabatic eBE in Table \ref{ebe} is 0.03 eV above
the excitation energy, the basis set convergence in Table \ref{basis} is
consistent with dipole-bound states, and the geometry of the radical in Table
\ref{ebe} is comparable to that of the optimized $2\ ^1A'$ state.

For CH$_2$SN, both the radical and the anion are planar while its first row
analogue, CH$_2$ON, is asymmetric in both electronic occupations.  The
``reluctance to hybridize" \citep{Kutzelnigg84, Driess96} in the sulfur atom
allows CH$_2$SN$^-$ to remain planar, whereas the MOs in CH$_2$ON$^-$ force the
hydrogens in the methylene group to pyramidalize.  This same behavior is
present in CH$_2$CN$^-$.  Hence, the geometrical differences and the subsequent
subtle changes to the orbitals for the inclusion of larger atoms from the
second row alone do not bring about valence states.  Otherwise, the
sulfur-containing anion, CH$_2$SN$^-$, would have at least one valence state.
While it is fascinating that these anions containing higher row atoms possess
dipole-bound excited states and the analogous first row anions do not, it is
the search for valence states that is driving this study.  Even though the
CH$_2$XM$^-$ anions are as close as possible to the structure of CH$_2$SiN$^-$,
these anions with second row elements do not give similar valence excitations.
However, the inclusion of a single second row atom systematically increases the
eBEs as compared to the first row only anions.  This increase appears to open
up the possibility for excited states to be present in the spectra of anions
containing second row atoms, but it does not guarantee the existence of
stable valence states. 





\subsection{Excited States of CCXH$_n^-$ Anions}

Molecules containing a carbide group have gained recent interest in
relation to the diffuse interstellar bands, \citep{Linnartz10, Maier11} and this
moiety is present in CCSiN$^-$, which has been shown to possess two valence
states \citep{Fortenberry112dbs}. As a result, simple anions of the CCXH$_n^-$
family where X=C, N, and O, as well as their second row analogues Si, S, and P,
are examined here.  An interesting feature of the entire CCXH$_n^-$ family of
anions is that the corresponding radicals all have very large dipole moments
from 3.409 D for C$_3$H to nearly 6 D for $C_{2v}$ CCNH$_2$.  This
behavior does not vary for the inclusion of the first or second row atoms; the
carbide moiety is responsible for the large dipole moments.
Previous computations of the dipole moment for CCOH are
consistent, \citep{Yamaguchi98, Fortenberry11dbs} and the 3.409 D dipole moment
of C$_3$H is also in line with previous CASSCF computations of its dipole
moment, 3.163 D, by Takahashi and Yamashita \citep{Takahashi96}.

The C$_3$H$^-$ system has received much attention over the past two
decades \citep{Takahashi96, Ochsenfeld97, Pachkov01, Pino04, Sheehan08} mainly
examining the anion to neutral (triplet to doublet) photochemistry.  Even
though it has been reported that the cyclic singlet is the most stable isomer
of C$_3$H$^-$, \citep{Sheehan08} consistency for the scope of this study and how
C$_3$H relates to the other anions from groups 15 and 16 in these CCXH$_n^-$
model systems mandates exploration of the excited state properties of the $C_s$
pseudo-linear $^1A'$ C$_3$H$^-$ and the silicon analogue, CCSiH$^-$.
CCSD(T)/aug-cc-pVTZ computations put the $^1A_1$ $c$-C$_3$H$^-$ isomer only 6.6
kcal/mol lower in energy than the $^1A'$ C$_3$H$^-$ isomer.  The pseudo-linear
singlet isomer is 6.9 kcal/mol lower in energy than $^3A'$ C$_3$H$^-$.

\begin{figure} 
    \includegraphics[width = 3.5 in]{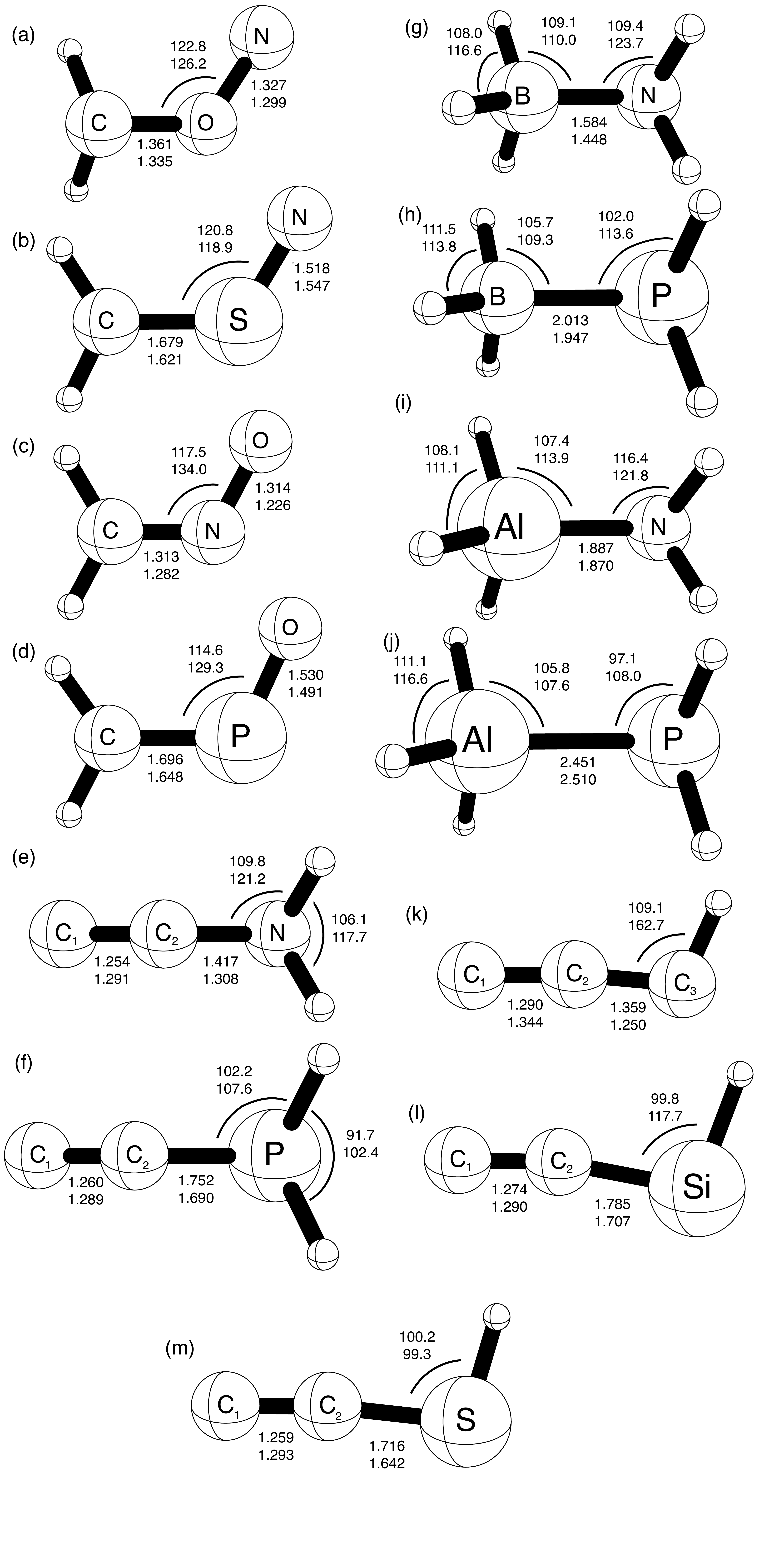}

\caption{CCSD(T)/aug-cc-pVTZ optimized structures of the ground states of the
closed-shell anions (top values; in units of \AA\ and degrees) and neutral
radicals (bottom values) of: (a) 1 $^1A$ CH$_2$ON$^-$ \& approximate (see text)
1 $^2A$ CH$_2$ON; (b) 1 $^1A'$ CH$_2$SN$^-$ \& 1 $^2A'$ CH$_2$SN; (c) 1 $^1A'$
CH$_2$NO$^-$ \& 1 $^2A'$ CH$_2$NO; (d) 1 $^1A'$ CH$_2$PO$^-$ \& 1 $^2A'$
CH$_2$PO; (e) 1 $^1A'$ CCNH$_2^-$ ($\angle$CCN$ =177.8^{\circ}$) \& 1 $^2B_1$
CCNH$_2$; (f) 1 $^1 A'$ CCPH$_2^-$ ($\angle$CCP$ =171.1^{\circ}$) \& 1 $^2A'$
CCPH$_2$ ($\angle$CCP$ =175.6^{\circ}$); (g) 1 $^1 A'$ BH$_3$NH$_2^-$ \& 1
$^2A''$ BH$_3$NH$_2$; (h) 1 $^1 A'$ BH$_3$PH$_2^-$ \& 1 $^2A'$ BH$_3$PH$_2$;
(i) 1 $^1 A'$ AlH$_3$NH$_2^-$ \& 1 $^2A'$ AlH$_3$NH$_2$; (j) 1 $^1 A'$
AlH$_3$PH$_2^-$ \& 1 $^2A'$ AlH$_3$PH$_2$; (k) 1 $^1A'$ C$_3$H$^-$
($\angle$CCC$ =174.2^{\circ}$) \& 1 $^2A'$ C$_3$H ($\angle$CCC$
=175.7^{\circ}$); (l) 1 $^1A'$ CCSiH$^-$ ($\angle$CCSi$ =170.2^{\circ}$) \& 1
$^2A'$ CCSiH ($\angle$CCSi$ =161.4^{\circ}$); and (m) 1 $^1A'$ CCSH$^-$
($\angle$CCS$ =173.9^{\circ}$) \& 1 $^2A''$ CCSH ($\angle$CCS$
=172.7^{\circ}$). }

\label{ebeFig}
\end{figure}

\begin{figure}
    \includegraphics[width = 3.5 in]{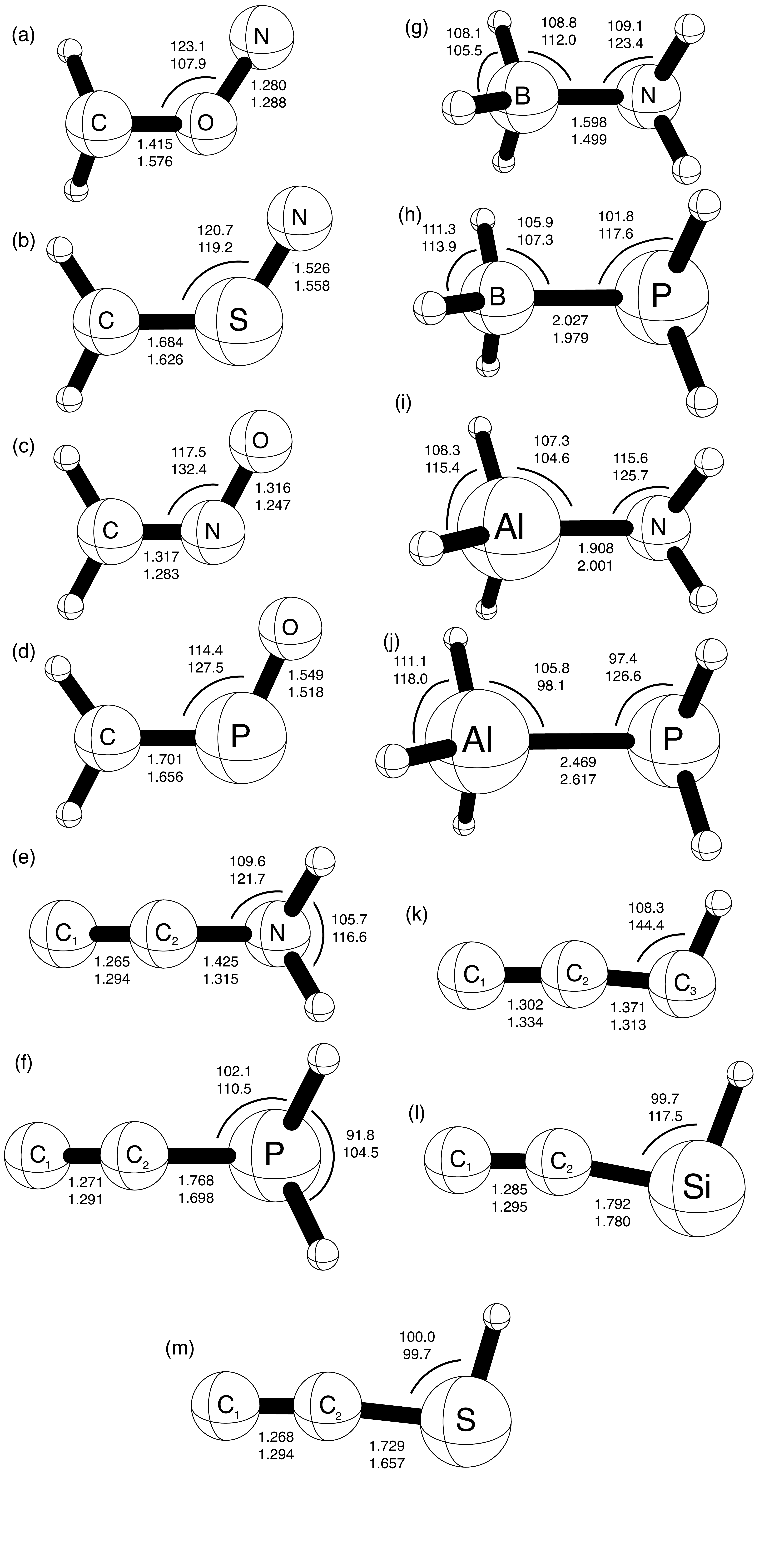}

\caption{CCSD/d-aug-cc-pVDZ optimized structures of ground (top values; again
in \AA\ and degrees) and first excited states (bottom values) of: (a) 1 $^1A$
\& 2 $^1A$ CH$_2$ON$^-$; (b) 1 $^1A'$ \& 2 $^1A'$ CH$_2$SN$^-$; (c) 1 $^1A'$ \&
2 $^1A'$ CH$_2$NO$^-$; (d) 1 $^1A'$ \& 2 $^1A'$ CH$_2$PO$^-$; (e) 1 $^1A'$
CCNH$_2^-$ ($\angle$CCN$ =177.9^{\circ}$) \& 1 $^1B_1$ CCNH$_2^-$; (f) 1 $^1
A'$ ($\angle$CCP$ =170.5^{\circ}$) \& 2 $^1A'$ CCPH$_2^-$ ($\angle$CCP$
=169.4^{\circ}$); (g) 1 $^1 A'$ \& 2 $^1A'$ BH$_3$NH$_2^-$; (h) 1 $^1 A'$ \& 2
$^1A'$ BH$_3$PH$_2^-$; (i) 1 $^1 A'$ \& 2 $^1A'$ AlH$_3$NH$_2^-$; (j) 1 $^1 A'$
\& 2 $^1A'$ AlH$_3$PH$_2^-$; (k) 1 $^1A'$ C$_3$H$^-$ ($\angle$CCC$
=174.0^{\circ}$) \& 1 $^1A''$ C$_3$H$^-$ ($\angle$CCC$ =174.0^{\circ}$); (l) 1
$^1A'$ CCSiH$^-$ ($\angle$CCSi$ =169.4^{\circ}$) \& 1 $^1A''$ CCSiH$^-$
($\angle$CCSi$ =170.5^{\circ}$); and (m) 1 $^1A'$ CCSH$^-$ ($\angle$CCS$
=173.3^{\circ}$) \& approximate (see text) 1 $^1A''$ CCSH$^-$ ($\angle$CCS$
=173.6^{\circ}$).}

\label{AdFig}
\end{figure}

Fascinatingly, pseudo-linear C$_3$H$^-$ appears to possess a valence state.
Table \ref{ebe} shows that the adiabatic excitation energy of the $1\ ^1A''
\leftarrow 1\ ^1A'$ transition at 0.93 eV is much less than the 1.83 eV
adiabatic eBE.  The vertical excitation energies for the increasingly diffuse
basis sets given in Table \ref{basis} decrease from 1.21 eV for cc-pVDZ merely
to 1.15 eV for aug-cc-pVDZ and to 1.14 eV for the converged excitation energy
for d- and t-aug-cc-pVDZ.  The converged vertical excitation energy is well
below the 2.34 vertical eBE.  Finally, the excited state geometry given in
Figure \ref{AdFig}k is markedly different from the radical geometry in Figure
\ref{ebeFig}k, especially for the H$-$C$_3-$C$_2$ bond angle.  Differently, the
vertical excitation energy of the $2\ ^1A' \leftarrow 1\ ^1A'$ transition at
2.37 eV for t-aug-cc-pVDZ is just above (but within computational accuracy of)
the vertical eBE.  However, the adiabatic computations put this state at 2.05
eV, 0.22 eV above the adiabatic eBE.  Our analysis strongly indicates that a
valence state is present in this anion.  It may yet possess a dipole-bound
state in addition to its valence excited state, but our computations do not
strongly support this interpretation.

The silicon analogue, CCSiH$^-$, exhibits similar behavior.  The 2.06 eV $1\
^1A'' \leftarrow 1\ ^1A'$ excitation energy is also well under the 3.11 eV
adiabatic eBE.  Table \ref{basis} also confirms that the $1\ ^1A''$ state is
valence in nature since the more diffuse basis sets lower the vertical
excitation energy for this state by only 0.04 eV.  The vertical eBE at 3.27 eV
is, again, well above the vertical excitation energy.  Differently, the $2\
^1A'$ state appears to be the dipole-bound state of CCSiH$^-$.  Like the
analogous $2\ ^1A'$ state of C$_3$H$^-$, this state is also more than 0.1 eV
above the eBE adiabatically,  but it is only 0.12 eV higher than the eBE here.
The EOM-CCSD/t-aug-cc-pVDZ vertical excitation energy of 3.26 eV is actually
0.01 eV less than the vertical eBE.  Hence, the dipole-bound state is more
likely to be present in the spectrum of CCSiH$^-$ than that of C$_3$H$^-$, but
the presence of dipole-bound states for these systems cannot be confirmed.
Regardless, both C$_3$H$^-$ and CCSiH$^-$ show clear indications of possessing
$1\ ^1A''$ valence states, the first time our methodology is reporting a
valence state for a small anion without silicon.


The valence state for C$_3$H$^-$ is probably the result of a more favorable
orbital arrangement arising from the presence of the carbene.  This allows the
valence $a''$ virtual orbital involved in the excitation to be more easily
accessed photochemically.  Figure \ref{ebeFig}k \& Figure \ref{ebeFig}l show
that the ground state structures of C$_3$H$^-$ and CCSiH$^-$ are both
substantially bent at the C$-$X$-$H (X=C, Si) bond angle.  The C$_3$H$^-$
ground state C$-$C$-$H bond angle is 109.1$^{\circ}$, while this same angle for
the radical in Figure \ref{ebeFig}k is 162.7$^{\circ}$.  The ground state
anionic C$-$Si$-$H bond angle is much closer to perpendicular at 99.8$^{\circ}$
as shown in Figure \ref{ebeFig}l.  The similar C$-$C$-$H bend in C$_3$H$^-$
allows its orbitals to behave more like those in CCSiH$^-$ than in previous
first-row-only anions studied.  One interpretation of this result is that the
central carbene carbon in the C$-$C$-$H bond angle in C$_3$H$^-$ has less
``hybridized" character since the C$-$C$-$H bond angle is noticeably less than
the desired 120$^{\circ}$ typically associated with an sp$^2$ carbon.  Such a
result is, again, a well-known property of singlet carbenes \citep{Bourissou00}.
Consequently, the carbene carbon is behaving more like silicon, which is much
less likely to hybridize \citep{Kutzelnigg84, Driess96} as is evidenced by its
even smaller C$-$Si$-$H bond angle in CCSiH$^-$ (99.8$^{\circ}$ in the ground
state) consistent with the 92.1$^{\circ}$ H$-$Si$-$H bond angle in singlet
SiH$_2$ \citep{Driess96, Dubois68}.  Further interpretation of the presence of
valence states and the C$-$X$-$H bond angles of less than 120$^{\circ}$ in both
C$_3$H$^-$ and CCSiH$^-$ is that the the amount of recoupled pair bonding in
C$_3$H$^-$ is much less than that present in typical carbon systems and is more
like that from higher row, early $p$-block atoms such as silicon \citep{Woon11,
Shaik12}. 

The presence of the CCOH$^-$ dipole-bound 1 $^1A''$ excited state is
unclear \citep{Fortenberry11dbs}.  However, the existence of the dipole-bound 1
$^1A''$ excited state of the sulfur analogue analyzed in this study, CCSH$^-$,
is more promising.  This corresponding radical was first analyzed
photochemically to compare its chemistry to CCOH and HCCO \citep{Venk91}. CCSH
shares similarities with CCOH in that the 4.492 D dipole moment is similar to
the 4.401 D dipole moment of CCOH.  The adiabatic excitation energy for the 1
$^1A''$ state of CCSH$^-$ is approximated to be 2.82 eV since the geometry
optimizations could not converge the root mean squared force to better than
$10^{-3} E_h/a_0$.  However, this approximate excitation energy is still 0.04
eV lower than the adiabatic eBE.  Additionally, the trend of vertical
excitations places this excitation beneath the vertical eBE, giving strong
evidence that this state is present in the electronic spectrum of CCSH$^-$.
Combining these factors with the similarities in the optimized geometry of the
radical and the approximate geometry of the anion, especially for the
H$-$S$-$C$_2$ bond angle from Figure \ref{ebeFig}m \& Figure \ref{AdFig}m, the
1 $^1A''$ state of CCSH$^-$ is a dipole-bound state.  However, no evidence is
present for a valence state of CCSH$^-$.

It was hoped that the large 5.903 D dipole moment of the stable \citep{Tao05}
CCNH$_2$ radical would increase the anionic eBE enough so that a valence state
could exist, but our computations show that no valence state is present for
this anion.  A dipole-bound state does appear to be present for CCNH$_2^-$, an
isomer of CH$_2$CN$^-$, which is the anion hypothesized to be a DIB
carrier \citep{Cordiner07, Fortenberry13CH2CN-}. The adiabatic excitation energy
of the 1 $^1B_1$ state (2 $^1A'$ in the vertical computations) of CCNH$_2^-$ is
2.00 eV, while the adiabatic eBE is nearly coincident at 1.97 eV, which is
within the computational limit for this state to exist.  The vertical
computations fully place the eBE higher than the excitation energy.
Additionally, the optimized bond angles and bond lengths of the anion excited
state and those of the radical given in Figure \ref{ebeFig}e \& Figure
\ref{AdFig}e show a strong correspondence, and the excited state is of $C_{2v}$
symmetry like that of the radical.  Hence, the first excited state of
CCNH$_2^-$ is the dipole-bound state.  No valence states exist.

The second row analogue, CCPH$_2^-$, exhibits similar dipole-bound behavior in
its 2 $^1A'$ state.  The dipole moment of the radical is smaller than the N
analogue, but it is still quite large at 4.759 D.  The 2 $^1A'$ state excites
at 3.16 eV from adiabatic computations.  The adiabatic eBE is slightly more
than this at 3.21 eV, which is strong evidence that this state is bound upon
excitation.  Additionally, the energy progression with increased diffuse
character of the basis set for the vertical excitation energy of this
CCPH$_2^-$ state clearly shows that it is indeed dipole-bound in nature and
also underneath the vertical eBE.  Comparison of the geometries in Figure
\ref{ebeFig}f \& Figure \ref{AdFig}f gives further evidence for the 2 $^1A'$
state to be dipole-bound.  Like its first row analogue, CCPH$_2^-$ possesses
one state that exhibits dipole-bound character, but no valence states appear to
be present even though the adiabatic eBE for CCPH$_2^-$ is 1.24 eV greater than
that of CCNH$_2^-$.

\subsection{Excited States of XH$_3$YH$_2^-$ Anions}

The XH$_3$YH$_2$ systems are all depicted in Figure \ref{ebeFig} and Figure
\ref{AdFig} in items g-j.  These systems are based on the dative bonding
molecule ammonia borane, BH$_3$NH$_3$, but the removal of a hydrogen from the
nitrogen changes the bonding chemistry.  For instance, experiment and theory
both agree that the B$-$N bond length in standard ammonia borane is 1.657
\AA\ \citep{Thorne83, Sams12}.  The lone pair on the ammonia molecule fills the
empty out-of-plane $p$ orbital in the borane.  Hence, the bonding environment
is not truly covalent but dative as evidenced by the long B$-$N bond length.
However, our systems have shorter B$-$N bond lengths than other dative
structures.  As shown in Figure \ref{ebeFig}g, the 1.448 \AA\ B$-$N bond length
in BH$_3$NH$_2$ is much less than the standard ammonia borane dative bond.
Even though the anion has a longer B$-$N bond than the radical at 1.584 \AA,
this bond is still much shorter than dative bonds in similar
systems \citep{Reinemann11}.

The molecules examined for this study appear to have a combination of dative
and covalent bonding due to the presence of bonds longer than typical covalent
bonds but not as long as dative bonds.  For instance, BH$_3$NH$_2^-$ can 
be thought of as the aminoborane anion, a $C_s$ molecule with the plane of
symmetry contained in the B$-$N bond bisecting the H$-$N$-$H bond angle.  The
borane still has its empty $p$ orbital which is filled not by the nitrogen's
lone pair but by another pair of electrons from the nitrogen giving a standard
covalent interaction while nitrogen retains its lone pair.  Hence, the bonding
is mixed from the covalency of the amino group and the dative interaction
originating from the borane.  The ground state of the BH$_3$NH$_2$ radical
actually has a different geometrical conformation than the anion.  Its geometry
is rotated so that the plane of symmetry in this $C_s$ radical contains both
the B$-$N bond and the H$-$N$-$H bond angle.

The aminoborane anion, BH$_3$NH$_2^-$, gives some evidence for the possession
of a dipole-bound excited state.  The radical has a dipole moment (2.524 D)
large enough for one to exist, but the 2.28 eV adiabatic eBE given in Table
\ref{ebe} is much lower than the adiabatic excitation energy to the 2 $^1A'$
state at 2.53 eV.  Differently, the vertical eBE of 2.88 eV given in Table
\ref{basis} is coincident with the converged vertical excitation energy for
this state.  Even though the energy difference between the eBE and excitation
energy adiabatically appears to be too great to support a dipole-bound state,
the vertical computations show that if this excited state does exist it is
probably dipole-bound.  The 2 $^1A'$ state of BH$_3$NH$_2^-$ has a geometry
similar in many respects to that of the radical from the the bottom values of
Figure \ref{ebeFig}g \& Figure \ref{AdFig}g, but the H$-$B$-$H bond angle
discrepancies highlight that these are geometrically different enough for the 2
$^1A'$ state not to be purely dipole-bound.  There is mixing in the character
of the excited state wavefunction, and it is great enough to stabilize this
state before it assumes an optimized geometry similar to that of the radical.
Regardless, our adiabatic energy difference in the eBE and excitation energy of
0.25 eV is probably too great for this excited state to exist.


Not surprisingly, both the BH$_3$PH$_2$ radical and anion have longer central
bonds than the analogous aminoborane compounds due to the larger valence
orbitals present in the phosphorus atom.  Figure \ref{ebeFig}h shows this
increase to be on the order of 0.4 \AA\ for the anion and 0.5 \AA\ for the
radical.  Also, the B$-$P$-$H bond angles are smaller in the BH$_3$PH$_2$
systems than the B$-$N$-$H bond angles in the BH$_3$NH$_2$ systems.  This is,
again, typical of second row atoms \citep{Driess96}.  Additionally, the $C_s$
BH$_3$PH$_2$ radical has its plane of symmetry bisecting the H$-$P$-$H bond
angle.  This is consistent with the corresponding anion but different from the
BH$_3$NH$_2$ radical.  The dipole moment of BH$_3$PH$_2$ is slightly larger
than its first row analogue at 2.889 D.  This makes it large enough for the
corresponding anion to possess a dipole-bound state, which it appears to have.
Table \ref{ebe} shows the adiabatic excitation to the 2 $^1A'$ state of
BH$_3$PH$_2^-$ to be 2.78 eV, while the adiabatic eBE is within our 0.1 eV
cutoff at 2.74 eV.  The adiabatic eBE is also nearly 0.5 eV greater than the
BH$_3$NH$_2^-$ adiabatic eBE.  The vertical excitation energy converges to the
same value as the vertical eBE, 3.26 eV, for the double-zeta basis sets.
Hence, the lone stable excited state, the 2 $^1A'$ state, of BH$_3$PH$_2^-$
appears to dipole-bound.


The insertion of aluminum into the XH$_3$YH$_2^-$ systems brings about some
interesting chemistry.  From Figure \ref{ebeFig}, the Al$-$N bond (1.887 \AA)
in AlH$_3$NH$_2^-$ is longer than the original B$-$N bond (1.584 \AA), but it
is not quite as long the B$-$P bond in BH$_3$PH$_2^-$ (2.013 \AA).
Additionally, the Al$-$N bonds in both the AlH$_3$NH$_2$ radical and anion are
shorter than the previously computed Al$-$N bond lengths in AlH$_3$NH$_3$,
which are more than 2.0 \AA, irrespective of the level of theory
employed \citep{Marsh92, Leboeuf95}. This is similar to the mixed covalent/dative
bonding in the boranes above.  However, the N$-$Al$-$H bond angle contained
within the plane of symmetry in these $C_s$ molecules is beyond perpendicular
for both the AlH$_3$NH$_2$ (68.2$^{\circ}$) and AlH$_3$PH$_2$ (69.7$^{\circ}$)
radicals.  The anions revert back to more typical bond angles \citep{Marsh92,
Leboeuf95} once the out of plane $p$ orbital in the N or P atom is filled to
give the standard lone pair.

Both alane species, AlH$_3$NH$_2^-$ and AlH$_3$PH$_2^-$, have two excited
states adiabatically either below or within 0.1 eV of their respective eBEs as
given in Table \ref{ebe}.  The inclusion of aluminum significantly increases
the eBEs as compared to the aminoborane or phosphinoborane anions, especially
when phosphorus is also included.  The corresponding radical to AlH$_3$NH$_2^-$
has the stronger dipole moment at 2.579 D whereas the dipole moment of
AlH$_3$PH$_2$ is somewhat below Gutsev and Adamowicz's \citep{Gutsev95a} 2.5 D
limit at 2.167 D.  This ensures that neither dipole moment is strong enough to
bind a second dipole-bound state and maybe not a first \citep{Fermi47,
Gutsev95a, Simons08}.  The only way that both states in each alane anion may be
retained is if one is valence.  Table \ref{basis} gives the vertical excitation
energies and eBEs for both the 2 $^1A'$ and 1 $^1A''$ states of both alanes.
In each of the two cases, the higher 1 $^1A''$ state excitation energy (3.85 eV
and 3.86 eV, respective of AlH$_3$NH$_2^-$ and AlH$_3$PH$_2^-$) is in the range
of what we expect for a classic dipole-bound state since the eBEs are 3.80 eV
and 3.85 eV, respectively.  The basis set convergence confirms the dipole-bound
character for both states of both anions.  However, if the lower energy excited
states are dipole-bound, only the lower energy states can exist since only one
dipole-bound state is allowed for these dipole moment magnitudes \citep{Fermi47}.

The classification of the lower 2 $^1A'$ states for each alane anion is a bit
more nebulous.  Due to the orbitals involved in the excitation with the highly
diffuse basis sets, no vertical excitation energies for the cc-pVDZ or cc-pVTZ
basis sets could be conclusively linked to comparable excitations utilizing the
more diffuse basis sets.  This is marked in Table \ref{basis}.  The lack of
correlation in vertical excitation energies does not indicate valence
excitation since all previous valence excitations characterized with this
methodology have had little change in the excitation energy from the cc-pVDZ
vertical excitation energy out to those computed with more diffuse basis sets.
Additionally, the vertical excitation energy convergence for the double zeta
basis sets from the aug-cc-pVDZ to the t-aug-cc-pVDZ level has previously been
less than 0.1 eV for valence states.  The change in vertical excitation
energies due to basis set augmentation for the 2 $^1A'$ state of
AlH$_3$NH$_2^-$ is 0.30 eV and 0.42 eV for the 2 $^1A'$ state of
AlH$_3$PH$_2^-$.  However, the change from d-aug-cc-pVDZ to t-aug-cc-pVDZ in
the aminoalane anion is only 0.01 eV and 0.04 eV in the phosphinoalane.  To
further complicate the interpretation, the optimized geometries of the 2 $^1A'$
states of both anions (given in Figures \ref{AdFig}i \& j) are not as similar
to the optimized geometries of the 1 $^2A'$ states of the radicals as one would
expect for dipole-bound states.  For instance, the Al$-$Y (Y=N \& P) bond
lengths are on the order of 0.1 \AA\ longer in the anion excited states than
they are in the corresponding radicals.  The Al$-$Y$-$H bond angles are larger
in the excited 2 $^1A'$ states of the anion than they are in the radical:
121.8$^{\circ}$ in the AlH$_3$NH$_2$ radical as opposed to 125.7$^{\circ}$ for
the 2 $^1A'$ state of the anion and 108.0$^{\circ}$ for the AlH$_3$PH$_2$
radical while the excited state of the anion has this bond angle at
126.6$^{\circ}$.  Similar to BH$_3$NH$_2$, these two closely related cases
appear to have mixed valence/dipole-bound character, the most of which for
any anionic excited states examined so far with this methodology.  It is
unclear if the valence character mixed into these excited states is large
enough to allow the higher, clearly dipole-bound states to exist above them.
Regardless, the presence of at least one excited state, the 2 $^1A'$, in the
electronic spectra is possible for both AlH$_3$NH$_2^-$ and AlH$_3$PH$_2^-$.

%
%

\section{Conclusions}

In the further examination of singlet excited states of small, closed-shell
molecular anions, specifically those that may possess valence excited states,
we have examined new species containing the second row atoms Al, Si, S, and P.
Of the twelve new anions examined here with further inclusion of CH$_2$NO$^-$
and CCOH$^-$ previously studied by \cite{Fortenberry11dbs}, nine exhibit
definite bound excited state character: CH$_2$SN$^-$, C$_3$H$^-$, CCSiH$^-$,
CCSH$^-$, CCNH$_2^-$, CCPH$_2^-$, BH$_3$PH$_2^-$, AlH$_3$NH$_2^-$, and
AlH$_3$PH$_2^-$.  Only two of these, C$_3$H$^-$ and CCNH$_2^-$, do not contain
second row atoms.  Adiabatic computations indicate that CH$_2$ON$^-$,
CH$_2$NO$^-$, CCOH$^-$, and BH$_3$NH$_2^-$ do not possess excited states,
dipole-bound or otherwise, and CH$_2$PO$^-$ likely does not either.

The only silicon system examined in this study, CCSiH$^-$, possesses two
excited states: one valence and one dipole-bound.  Interestingly, the carbon
analogue, C$_3$H$^-$, also exhibits both a single valence excited state and a
dipole-bound excited state.  This is possibly the result of similar orbital
arrangements present in both systems that was not present in those first and
second row analogues previously studied.  In this regard, silicon does not
appear unique in its ability to foster valence singlet excited states in
anions.  Even so, when varying the atoms at all non-hydrogen and
non-dipole-inducing positions, only the inclusion of group 14 atoms has been
shown to contribute to stable valence excitations for closed-shell anions.
This could also be true for inclusion of a directed group 15 and 16 atom since
the presence of these second row atoms increases the eBEs noticeably as
compared to their first row analogues.  However, the properties of anions with
group 15 and 16 atoms do not appear as favorable for the existence of valence
excited states of closed-shell anions based on the systems analyzed both here
and previously \citep{Fortenberry11dbs, Fortenberry112dbs}.

Mixing of valence and dipole-bound character in the excited state wavefunctions
can be more substantial than originally thought.  This is present in some of
the carbide anions, but the aminoborane anion and its analogues, especially the
alanes, have substantial mixing of valence and dipole-bound excited state
character in their first excited states.  The vertical excitation energies for
AlH$_3$NH$_2^-$ and AlH$_3$PH$_2^-$, for example, shift noticeably for more
diffuse basis sets unlike typical valence excitations but not as much as the
shifts observed for established dipole-bound anions.  The mixing is also
showcased in the lack of correspondence between the optimized radical and
excited 2 $^1A'$ state geometries.  It is unclear if the amount of mixing in
these first excited states gives enough valence character for the second state
to play the role of the ``threshold resonance," but the order of magnitude
variance in the oscillator strengths (Table \ref{basis}) between the lower and
stronger 2 $^1A'$ states and the higher and less intense 1 $^1A''$ states of
each alane anion would make this experimentally measurable.


The presence of second row atoms in the types of closed-shell anions examined
thus far increases the likelihood of dipole-bound excited states since the eBE
and dipole moments are both increased with the larger atoms.  Combining these
facts with the increased prevalence of mixing for valence and dipole-bound
character in the excited state wavefunction indicates that the larger atoms
create environments more favorable for excitation of electrons in closed-shell
anions, but the anions built around group 14 atoms, especially silicon, are the
most favorable for anion excitation.

Finally, direct attribution of known DIBs to computed anionic excitations is
impractical for the present level of computational accuracy.  However, the
accuracy range combined with the density of the DIBs (with a band present every
nm or so between 400 nm and 880 nm) gives many options for correlation between
the computed transitions of our set of anions and the various
DIBs \citep{Joblin90, Jenniskens94, Jenn}.  In fact, one of the longest
wavelength DIBs at 1317.5 nm \citep{Joblin90} is in the proper range to be
related potentially to the 1328 nm dipole-bound $1\ ^1A'' \leftarrow 1\ ^1A'$
transition of C$_3$H$^-$, while the 604 nm valence $2\ ^1A'' \leftarrow 1\
^1A'$ transition is within 10 nm of more than a dozen known
DIBs \citep{Jenniskens94, Jenn}.

\section{Acknowledgements}

RCF was funded for this work partly by the NASA Postdoctoral Program
administered through Oak Ridge Associated Universities.  Funding for this work
has also come from the U.S. National Science Foundation: award CHE-1058420 and
a Multi-User Chemistry Research Instrumentation and Facility (CRIF:MU) award
CHE-0741927 and by the Virginia Space Grant Consortium in the form of a
Graduate Research Fellowship for RCF.  The figures were generated in part with
the CheMVP program made available by Dr.\ Andrew Simmonett of the University of
Georgia.  RCF would like to thank Dr.~David Woon of the University of Illinois
for discussions regarding the bonding of the higher row atoms and Dr.~Fabio
Carelli of Spienza$-$University of Rome for insights into the nature of
electronic excitations and binding in anions.  Dr.~Timothy J.~Lee of the NASA
Ames Research Center is also due thanks for his encouragement on finalizing
this project.  A tremendous debt of gratitude belongs to Prof.~T.~Danial
Crawford of Virginia Tech for providing the computer resources necessary to
execute the computations as well as for guiding many aspects of this research
and for assistance in editing the manuscript.

\bibliographystyle{apj}


\begin{thebibliography}{86}
\expandafter\ifx\csname natexlab\endcsname\relax\def\natexlab#1{#1}\fi

\bibitem[{CFO(2010)}]{CFOUR}
 2010, {\sc CFOUR}, a quantum chemical program package written by J.F. Stanton,
  J. Gauss, M.E. Harding, P.G. Szalay with contributions from A.A. Auer, R.J.
  Bartlett, U. Benedikt, C. Berger, D.E. Bernholdt, Y.J. Bomble, O.
  Christiansen, M. Heckert, O. Heun, C. Huber, T.-C. Jagau, D. Jonsson, J.
  Jus\'{e}lius, K. Klein, W.J. Lauderdale, D.A. Matthews, T. Metzroth, D.P.
  O'Neill, D.R. Price, E. Prochnow, K. Ruud, F. Schiffmann, S. Stopkowicz, A.
  Tajti, J. V\'azquez, F. Wang, J.D. Watts and the integral packages {\sc
  MOLECULE} (J. Alml\"of and P.R. Taylor), {\sc PROPS} (P.R. Taylor), {\sc
  ABACUS} (T. Helgaker, H.J. Aa. Jensen, P. J{\o}rgensen, and J. Olsen), and
  {\sc ECP} routines by A. V. Mitin and C. van W\"ullen. For the current
  version, see {\tt http://www.cfour.de}.

\bibitem[{Ard {et~al.}(2009)Ard, Garrett, Compton, Adamowicz, \&
  Stepanian}]{Ard09}
Ard, S., Garrett, W.~R., Compton, R.~N., Adamowicz, L., \& Stepanian, S.~G.
  2009, Chem. Phys. Lett., 473, 223

\bibitem[{Binkley \& Thorne(1983)}]{Binkley83}
Binkley, J.~S., \& Thorne, L.~R. 1983, J. Chem. Phys., 79, 2932

\bibitem[{Bourissou {et~al.}(2000)Bourissou, Guerret, Gabba\"{i}, \&
  Bertrand}]{Bourissou00}
Bourissou, D., Guerret, O., Gabba\"{i}, F.~P., \& Bertrand, G. 2000, Chem.
  Rev., 100, 39

\bibitem[{Brinkman {et~al.}(1993)Brinkman, Gunther, Schafer, \&
  Brauman}]{Brinkman93}
Brinkman, E.~A., Gunther, E., Schafer, O., \& Brauman, J.~I. 1993, J. Chem.
  Phys., 100, 1840

\bibitem[{Brown \& Borden(2000)}]{Brown00}
Brown, E.~C., \& Borden, W.~T. 2000, Organometallics, 19, 2208

\bibitem[{Chen \& Chen(2001)}]{Chen01}
Chen, E.~S., \& Chen, E. C.~M. 2001, Biochem. Biophys. Res. Commun., 289, 421

\bibitem[{Chesnut(2002)}]{Chesnut02}
Chesnut, D.~B. 2002, Heteroat. Chem., 13, 53

\bibitem[{Cordiner \& Sarre(2007)}]{Cordiner07}
Cordiner, M.~A., \& Sarre, P.~J. 2007, Astron. Astrophys., 472, 537

\bibitem[{Coulson \& Walmsley(1967)}]{Coulson67}
Coulson, C.~A., \& Walmsley, M. 1967, Proc. Phys. Soc., 91, 31

\bibitem[{Crawford \& Dalgarno(1967)}]{Crawford67}
Crawford, O.~H., \& Dalgarno, A. 1967, Chem. Phys. Lett., 1, 23

\bibitem[{Crawford \& Garrett(1977)}]{Crawford77}
Crawford, O.~H., \& Garrett, W.~R. 1977, J. Chem. Phys., 66, 4968

\bibitem[{Crawford {et~al.}(2007)Crawford, Sherrill, Valeev, Fermann, King,
  Leininger, Brown, Janssen, Kenny, Seidl, \& Allen}]{Psi3}
Crawford, T.~D., Sherrill, C.~D., Valeev, E.~F., {et~al.} 2007, J. Comput.
  Chem., 28, 1610

\bibitem[{Driess \& Gr\"{u}tzmacher(1996)}]{Driess96}
Driess, M., \& Gr\"{u}tzmacher, H. 1996, Angew. Chem. Int. Ed. Engl., 35, 828

\bibitem[{Dubois(1968)}]{Dubois68}
Dubois, I. 1968, Can. J. Phys., 46, 2485

\bibitem[{Dunning(1989)}]{Dunning89}
Dunning, T.~H. 1989, J. Chem. Phys., 90, 1007

\bibitem[{Dunning {et~al.}(2001)Dunning, Peterson, \& Wilson}]{Dunning01}
Dunning, T.~H., Peterson, K.~A., \& Wilson, A.~K. 2001, J. Chem. Phys., 114,
  9244

\bibitem[{Fermi \& Teller(1947)}]{Fermi47}
Fermi, E., \& Teller, E. 1947, Phys. Rev., 72, 399

\bibitem[{Fortenberry \& Crawford(2011{\natexlab{a}})}]{Fortenberry112dbs}
Fortenberry, R.~C., \& Crawford, T.~D. 2011{\natexlab{a}}, J. Phys. Chem. A,
  115, 8119

\bibitem[{Fortenberry \& Crawford(2011{\natexlab{b}})}]{Fortenberry11dbs}
---. 2011{\natexlab{b}}, J. Chem. Phys., 134, 154304

\bibitem[{Fortenberry {et~al.}(2013)Fortenberry, Crawford, \&
  Lee}]{Fortenberry13CH2CN-}
Fortenberry, R.~C., Crawford, T.~D., \& Lee, T.~J. 2013, Astrophys. J., 762,
  121

\bibitem[{Gauss {et~al.}(1991)Gauss, Stanton, \& Bartlett}]{Gauss-UHF91}
Gauss, J., Stanton, J.~F., \& Bartlett, R.~J. 1991, J. Chem. Phys., 95, 2623

\bibitem[{Gutowski {et~al.}(1996)Gutowski, Skurksi, Boldyrev, Simons, \&
  Jordan}]{Gutowski96}
Gutowski, M., Skurksi, P., Boldyrev, A.~I., Simons, J., \& Jordan, K.~D. 1996,
  Phys. Rev., 54, 1906

\bibitem[{Gutsev \& Adamowicz(1995)}]{Gutsev95a}
Gutsev, G., \& Adamowicz, A. 1995, Chem. Phys. Lett., 235, 377

\bibitem[{Helgaker {et~al.}(2004)Helgaker, Ruden, J{\o}rgensen, Olsen, \&
  Klopper}]{Helgaker04}
Helgaker, T., Ruden, T.~A., J{\o}rgensen, P., Olsen, J., \& Klopper, W. 2004,
  J. Phys. Org. Chem., 17, 913

\bibitem[{Inostroza \& Senent(2010)}]{Inostroza10}
Inostroza, N., \& Senent, M.~L. 2010, J. Chem. Phys., 133, 184107

\bibitem[{Jenniskens \& Desert(1994)}]{Jenniskens94}
Jenniskens, P., \& Desert, F.-X. 1994, Astron. Astrophys. Supp. Ser., 106, 39

\bibitem[{Jenniskins \& D\'{e}sert(1995)}]{Jenn}
Jenniskins, P., \& D\'{e}sert, F.-X. 1995, in The Diffuse Interstellar Bands
  (Dordrecht, Netherlands: Kluwer), 39--52

\bibitem[{Joblin {et~al.}(1990)Joblin, Maillard, D'Hendecourt, \&
  L\'{e}ger}]{Joblin90}
Joblin, C., Maillard, J.~P., D'Hendecourt, L., \& L\'{e}ger, A. 1990, Nature,
  346, 729

\bibitem[{Jochnowitz \& Maier(2008)}]{Jochnowitz08}
Jochnowitz, E.~B., \& Maier, J.~P. 2008, Ann. Rev. Phys. Chem., 59, 519

\bibitem[{Jordan \& Luken(1976)}]{Jordan76}
Jordan, K.~D., \& Luken, W. 1976, J. Chem. Phys., 64, 2760

\bibitem[{Jordan \& Wang(2003)}]{Jordan03}
Jordan, K.~D., \& Wang, F. 2003, Ann. Rev. Phys. Chem., 54, 367

\bibitem[{Kutzelnigg(1984)}]{Kutzelnigg84}
Kutzelnigg, W. 1984, Angew. Chem. Int. Ed. Engl., 23, 272

\bibitem[{Leboeuf {et~al.}(1995)Leboeuf, Russo, Salahub, \&
  Toscano}]{Leboeuf95}
Leboeuf, M., Russo, N., Salahub, D.~R., \& Toscano, M. 1995, J. Chem. Phys.,
  103, 7413

\bibitem[{Lee \& Rendell(1991)}]{Lee91}
Lee, T.~J., \& Rendell, A.~P. 1991, J. Chem. Phys., 94, 6229

\bibitem[{Lee \& Scuseria(1995)}]{Lee95Accu}
Lee, T.~J., \& Scuseria, G.~E. 1995, in Quantum Mechanical Electronic Structure
  Calculations with Chemical Accuracy, ed. S.~R. Langhoff (Dordrecht: Kluwer
  Academic Publishers), 47--108

\bibitem[{Li {et~al.}(1995)Li, {Van Zee}, Weltner, \& Raghavachari}]{Li95}
Li, S., {Van Zee}, R.~J., Weltner, W., \& Raghavachari, K. 1995, Chem. Phys.
  Lett., 243, 275

\bibitem[{Linnartz {et~al.}(2010)Linnartz, Wehres, Walker, Bohlender, Teilens,
  Motylewski, \& Maier}]{Linnartz10}
Linnartz, H., Wehres, N., Walker, G. A.~H., {et~al.} 2010, Astron. Astrophys.,
  511, L3

\bibitem[{Lykke {et~al.}(1987)Lykke, Neumark, Andersen, Trapa, \&
  Lineberger}]{Lykke87}
Lykke, K.~R., Neumark, D.~M., Andersen, T., Trapa, V.~J., \& Lineberger, W.~C.
  1987, J. Chem. Phys., 87, 6842

\bibitem[{Maier(1998)}]{Maier98}
Maier, J.~P. 1998, J. Phys. Chem., 102, 3462

\bibitem[{Maier {et~al.}(2011)Maier, Walker, Bohlender, Mazzotti, Raghunandan,
  Fulara, Garkusha, \& Nagy}]{Maier11}
Maier, J.~P., Walker, G. A.~H., Bohlender, D.~A., {et~al.} 2011, Astrophys. J.,
  726, 41

\bibitem[{Marsh {et~al.}(1992)Marsh, Hamilton, Xie, \& {Schaefer
  III}}]{Marsh92}
Marsh, C. M.~B., Hamilton, T.~P., Xie, Y., \& {Schaefer III}, H.~F. 1992, J.
  Chem. Phys., 96, 5310

\bibitem[{Mead {et~al.}(1984)Mead, Lykke, \& Lineberger}]{Mead84}
Mead, R.~D., Lykke, K.~R., \& Lineberger, W.~C. 1984, J. Chem. Phys., 81, 4883

\bibitem[{Meloni {et~al.}(2004)Meloni, Sheehan, Ferguson, \&
  Neumark}]{Meloni04}
Meloni, G., Sheehan, S.~M., Ferguson, M.~J., \& Neumark, D.~M. 2004, J. Phys.
  Chem. A., 108, 9750

\bibitem[{Monkhorst(1977)}]{Monkhorst77}
Monkhorst, H.~J. 1977, Int. J. Quantum Chem. Symp., 11, 421

\bibitem[{Mukherjee \& Mukherjee(1979)}]{Mukherjee79}
Mukherjee, D., \& Mukherjee, P.~K. 1979, Chem. Phys., 39, 325

\bibitem[{Mullin {et~al.}(1993)Mullin, Murray, Schulz, \&
  Lineberger}]{Mullin93}
Mullin, A.~S., Murray, K.~K., Schulz, C.~P., \& Lineberger, W.~C. 1993, J.
  Phys. Chem., 97, 10281

\bibitem[{Mullin {et~al.}(1992)Mullin, Murray, Schulz, Szaflarski, \&
  Lineberger}]{Mullin92}
Mullin, A.~S., Murray, K.~K., Schulz, C.~P., Szaflarski, D.~M., \& Lineberger,
  W.~C. 1992, Chem. Phys., 166, 207

\bibitem[{Ochsenfeld {et~al.}(1997)Ochsenfeld, Kaiser, Lee, Suits, \&
  Head-Gordon}]{Ochsenfeld97}
Ochsenfeld, C., Kaiser, R.~I., Lee, Y.~T., Suits, A.~G., \& Head-Gordon, M.
  1997, J. Chem. Phys., 106, 4141

\bibitem[{Owens {et~al.}(2006)Owens, Larkin, \& {Schaefer III}}]{Owens06}
Owens, Z.~T., Larkin, J.~D., \& {Schaefer III}, H.~F. 2006, J. Chem. Phys.,
  125, 164322

\bibitem[{Pachkov {et~al.}(2001)Pachkov, Pino, Tulej, \& Maier}]{Pachkov01}
Pachkov, M., Pino, T., Tulej, M., \& Maier, J.~P. 2001, Mol. Phys., 99, 1397

\bibitem[{Peterson \& Dunning(1995)}]{cc-pVXZ}
Peterson, K.~A., \& Dunning, T.~H. 1995, J. Chem. Phys., 102, 2032

\bibitem[{Pino {et~al.}(2004)Pino, Pachkov, Tulej, Xu, Jungen, \&
  Maier}]{Pino04}
Pino, T., Pachkov, M., Tulej, M., {et~al.} 2004, Mol. Phys., 102, 1881

\bibitem[{Power(1999)}]{Power99}
Power, P.~P. 1999, Chem. Rev., 99, 3463

\bibitem[{Raghavachari {et~al.}(1989)Raghavachari, Trucks., Pople, \&
  Head-Gordon}]{Ragh89}
Raghavachari, K., Trucks., G.~W., Pople, J.~A., \& Head-Gordon, M. 1989, Chem.
  Phys. Lett., 157, 479

\bibitem[{Reilly {et~al.}(2012)Reilly, Kokkin, Zhuang, Guptac, Nagarajan,
  Fortenberry, Maier, Steimle, Stanton, \& McCarthy}]{Reilly12}
Reilly, N.~J., Kokkin, D.~L., Zhuang, X., {et~al.} 2012, J. Chem. Phys., 136,
  194307

\bibitem[{Reinemann {et~al.}(2011)Reinemann, Wright, Wolfe, Tschumper, \&
  Hammer}]{Reinemann11}
Reinemann, D.~N., Wright, A.~M., Wolfe, J.~D., Tschumper, G.~S., \& Hammer,
  N.~I. 2011, J. Phys. Chem. A, 115, 6426

\bibitem[{Sams {et~al.}(2012)Sams, Xantheas, \& Blake}]{Sams12}
Sams, R.~L., Xantheas, S.~S., \& Blake, T.~A. 2012, J. Phys. Chem. A, 116, 3124

\bibitem[{Sarre(2000)}]{Sarre00}
Sarre, P.~J. 2000, Mon. Not. R. Astron. Soc., 313, L14

\bibitem[{Sarre(2006)}]{Sarre06}
---. 2006, J. Mol. Spectrosc., 238, 1

\bibitem[{Scheiner {et~al.}(1987)Scheiner, Scuseria, Rice, Lee, \& {Schaefer
  III}}]{ScheinerRHF87}
Scheiner, A.~C., Scuseria, G.~E., Rice, J.~E., Lee, T.~J., \& {Schaefer III},
  H.~F. 1987, J. Chem. Phys., 87, 5361

\bibitem[{Shaik {et~al.}(2012)Shaik, Danovitch, Wu, Su, Rzepa, \&
  Hiberty}]{Shaik12}
Shaik, S., Danovitch, D., Wu, W., {et~al.} 2012, Nat. Chem., 4, 195

\bibitem[{Shapley \& Bacsky(1998)}]{Shapley98}
Shapley, W.~A., \& Bacsky, G.~B. 1998, Theor. Chem. Acc., 100, 212

\bibitem[{Shapley \& Bacsky(1999)}]{Shapley99}
---. 1999, J. Phys. Chem. A, 103, 4505

\bibitem[{Shavitt \& Bartlett(2009)}]{Shavitt09}
Shavitt, I., \& Bartlett, R.~J. 2009, Many-Body Methods in Chemistry and
  Physics: MBPT and Coupled-Cluster Theory (Cambridge: Cambridge University
  Press)

\bibitem[{Sheehan {et~al.}(2008)Sheehan, Parsons, Zhou, Garand, Yen, Moore, \&
  Neumark}]{Sheehan08}
Sheehan, S.~M., Parsons, B.~F., Zhou, J., {et~al.} 2008, J. Chem. Phys., 128,
  034301

\bibitem[{Simons(2008)}]{Simons08}
Simons, J. 2008, J. Phys. Chem. A., 112, 6401

\bibitem[{Simons(2011)}]{Simons11}
---. 2011, Annu. Rev. Phys. Chem., 62, 107

\bibitem[{Skurski \& Gutowski(2000)}]{Skurski00TCNQ}
Skurski, P., \& Gutowski, M. 2000, J. Mol. Struct., 531, 339

\bibitem[{Skurski {et~al.}(2000)Skurski, Gutowski, \& Simons}]{Skurski00}
Skurski, P., Gutowski, M., \& Simons, J. 2000, Int. J. Quant. Chem., 80, 1024

\bibitem[{Sobczyk {et~al.}(2003)Sobczyk, Skurski, \& Simons}]{Sobczyk03}
Sobczyk, M., Skurski, P., \& Simons, J. 2003, J. Phys. Chem. A, 107, 7084

\bibitem[{Stanton \& Bartlett(1993)}]{Stanton93EOM}
Stanton, J.~F., \& Bartlett, R.~J. 1993, J. Chem. Phys., 98, 7029

\bibitem[{Stanton \& Gauss(1994)}]{Stanton94:EOMIP}
Stanton, J.~F., \& Gauss, J. 1994, J. Chem. Phys., 101, 8938

\bibitem[{Takahashi \& Yamashita(1996)}]{Takahashi96}
Takahashi, J., \& Yamashita, K. 1996, J. Chem. Phys., 104, 6613

\bibitem[{Tao(2005)}]{Tao05}
Tao, Y. 2005, J. Nat. Sci. Heilongjiang Univ., 22, 241

\bibitem[{Thorne {et~al.}(1983)Thorne, Suenram, \& Lovas}]{Thorne83}
Thorne, L.~R., Suenram, R.~D., \& Lovas, F.~J. 1983, J. Chem. Phys., 78, 167

\bibitem[{Turner(1977)}]{Turner77}
Turner, J.~E. 1977, Am. J. Phys., 45, 758

\bibitem[{Turner \& Fox(1966)}]{Turner66}
Turner, J.~E., \& Fox, K. 1966, Phys. Lett., 23, 547

\bibitem[{Venkatasubramanian \& Krishnamachari(1991)}]{Venk91}
Venkatasubramanian, R., \& Krishnamachari, S. L. N.~G. 1991, Indian J. Pure
  Appl. Phys., 29, 697

\bibitem[{Wang \& Jordan(2002)}]{Wang02}
Wang, F., \& Jordan, K.~D. 2002, J. Chem. Phys., 116, 6973

\bibitem[{Watts {et~al.}(1992)Watts, Gauss, \& Bartlett}]{Watts-UHF92}
Watts, J.~D., Gauss, J., \& Bartlett, R.~J. 1992, Chem. Phys. Lett., 200, 1

\bibitem[{Woon \& Dunning(2009)}]{Woon09SFn}
Woon, D.~E., \& Dunning, T.~H. 2009, J. Phys. Chem. A, 113, 7915

\bibitem[{Woon \& Dunning(2010)}]{Woon10PFn}
---. 2010, J. Phys. Chem. A, 114, 8845

\bibitem[{Woon \& Dunning(2011)}]{Woon11}
---. 2011, J. Comput. Theor. Chem., 963, 7

\bibitem[{Yamaguchi {et~al.}(1998)Yamaguchi, {Rienstra-Kiracofe}, Stephens, \&
  {Schaefer III}}]{Yamaguchi98}
Yamaguchi, Y., {Rienstra-Kiracofe}, J.~C., Stephens, J.~C., \& {Schaefer III},
  H.~F. 1998, Chem. Phys. Lett., 291, 509

\bibitem[{Zhoa {et~al.}(2009)Zhoa, Hou, Shu, Zhang, \& Sun}]{Zhao09}
Zhoa, Z.-X., Hou, C.-Y., Shu, X., Zhang, H.-X., \& Sun, C. 2009, Theor. Chem.
  Acc., 124, 85

\end{thebibliography}

\newpage

\renewcommand{\baselinestretch}{1}

\setcounter{table}{0}
\begin{table}
\caption{Dipole moments of the corresponding neutral radical (in Debye),
adiabatic electron binding energies (in eV), and singlet adiabatic
excited-state transition energies (in eV) and wavelengths (in nm) for several
second-row anions.}

\begin{tabular}{l c c l c l c c c}
\hline\hline
\label{ebe}

 & \multicolumn{2}{c}{\raisebox{-0.5ex}[0cm][0cm]{Radical Dipole Moment}}
& \hspace{0.05 in} & & \hspace{0.05 in} & &
\multicolumn{2}{c}{\raisebox{-0.5ex}[0cm][0cm]{Transition$^c$}}\\

\cline{2-3} \cline{8-9} 

Molecule \hspace{0.01 in} & This Work$^a$ & Previous & & eBE$^b$ & & Transition
& Energy & Wavelength \\

\hline\hline

CH$_2$ON$^-$ & 2.335 & $\ldots$ &  & 0.46$^d$ &  & $2\ ^1A \leftarrow 1\ ^1A$ &
0.89 & 1396 \\ 

\cline{7-9}

 &  &  &  &  &  & $3\ ^1A \leftarrow 1\ ^1A$ & 1.08 & 1149 \\ 

\hline

CH$_2$SN$^-$ & 2.703 & $\ldots$ &  & 1.98 &  & $2\ ^1A' \leftarrow 1\ ^1A'$ & 1.95
& 634 \\ 

\cline{7-9}

 &  &  &  &  &  & $1\ ^1A'' \leftarrow 1\ ^1A'$ & 2.14 & 580 \\ 

\hline

CH$_2$NO$^-$$^e$ & 2.317 & $\ldots$ &  & 1.42 & & $2\ ^1A'
\leftarrow 1\ ^1A'$ & 1.55 & 799 \\ 

\hline

CH$_2$PO$^-$ & 2.477 & $\ldots$ &  & 2.82 &  & $1\ ^1A'' \leftarrow 1\ ^1A'$ &
2.89 & 429 \\ 
 & & & & & & $2\ ^1A' \leftarrow 1\ ^1A'$ & 3.08 & 403 \\

\hline

C$_3$H$^-$ & 3.409 & 3.163$^f$ &  & 1.83 &  & $1\ ^1A'' \leftarrow 1\
^1A'$ & 0.93 & 1328 \\ 

\cline{7-9}

 &  &  &  &  &  & $2\ ^1A' \leftarrow 1\ ^1A'$ & 2.05 & 604 \\ 

\hline

CCSiH$^-$ &  & $\ldots$ &  & 3.11 &  & $1\ ^1A'' \leftarrow 1\ ^1A'$ & 2.06 &
602 \\ 

\cline{7-9}

 &  &  &  &  &  & $2\ ^1A' \leftarrow 1\ ^1A'$ & 3.23 & 384 \\ 

\hline

CCOH$^-$$^e$ & 4.401 & 4.410$^g$ &  & 2.52 &  & 1$\ ^1A''\leftarrow 1\ ^1A'$
& 2.43 & 511 \\

\hline

CCSH$^-$ & 4.492 & $\ldots$ &  & 2.86 &  & $1\ ^1A'' \leftarrow 1\ ^1A'$ &
2.82$^d$ & 440 \\ 

\hline

CCNH$_2^-$ & 5.903 & $\ldots$  &  & 1.97 &  & $1\ ^1B_1 \leftarrow 1\ ^1A'$ & 2.00 & 620 \\ 

\hline

CCPH$_2^-$ & 4.759 & $\ldots$ &  & 3.21 &  & $2\ ^1A' \leftarrow 1\ ^1A'$ &
3.16 & 392 \\ 

\hline

BH$_3$NH$_2^-$ & 2.524 & $\ldots$ &  & 2.28 &  & $2\ ^1A' \leftarrow 1\ ^1A'$ &
2.53 & 490 \\ 

\hline

BH$_3$PH$_2^-$ & 2.889 & $\ldots$ &  & 2.74 &  & $2\ ^1A' \leftarrow 1\ ^1A'$ &
2.78 & 446 \\ 

\hline

AlH$_3$NH$_2^-$ & 2.579 & $\ldots$ &  & 3.22 &  & $2\ ^1A' \leftarrow 1\ ^1A'$ &
3.17 & 391 \\

\cline{7-9}

 &  &  &  &  &  & $1\ ^1A'' \leftarrow 1\ ^1A'$ & 3.32 & 373 \\

\hline

AlH$_3$PH$_2^-$ & 2.167 & $\ldots$ &  & 3.25 &  & $2\ ^1A' \leftarrow 1\ ^1A'$ &
3.21 & 387 \\

\cline{7-9}

 &  &  &  &  &  & $1\ ^1A'' \leftarrow 1\ ^1A'$ & 3.35 & 370 \\ 

\hline
\end{tabular}

$^a$UHF-CCSD/aug-cc-pVTZ values for the radicals at the UHF-CCSD(T)/aug-cc-pVTZ geometries.
$^b$The differences between the RHF- (anion) and UHF- (radical) CCSD(T)/aug-cc-pVTZ energies.
$^c$Adiabatic EOM-CCSD/d-aug-cc-pVDZ values.
$^d$This value is from the non-minimum geometry but functions as an upper bound estimate.
$^e$Computed by \cite{Fortenberry11dbs}.
$^f$CASSCF/D95(d,p) computation from \cite{Takahashi96}.
$^g$CISD/TZ3P(2f,2d) result from \cite{Yamaguchi98}.

\end{table}

\setcounter{table}{1}
\begin{table}

\caption{EOM-CCSD vertical excitation energies (in eV), oscillator
strengths,$^a$ and vertical electron binding energies (in eV) from
ground state CCSD(T)/aug-cc-pVTZ geometries for several basis
sets.$^b$}

\scriptsize
\begin{tabular}{l c | c c c c | c c c | c || c}
\hline\hline
\label{basis}

Molecule & Transition & pVDZ & apVDZ & dapVDZ & tapVDZ & pVTZ & apVTZ
& dapVTZ & $f$ & eBE$^c$ \\

\hline\hline

CH$_2$ON$^-$ & $2\ ^1A \leftarrow 1\ ^1A$ & 3.82 &
1.82 & 1.17 & 1.03 & 3.54 & 1.90 & 1.37 & 1$\times 10^{-3}$ & 1.53 \\

\cline{2-10}

 & $3\ ^1A \leftarrow 1\ ^1A$ & 1.26 & 1.13 & 1.07 & 1.06 & 1.21 & 1.10 & 1.08
& 7$\times 10^{-3}$ & \\

\hline

CH$_2$SN$^-$ & $2\ ^1A' \leftarrow 1\ ^1A'$ & 5.95 &
3.18 & 2.09 & 1.95 & 5.81 & 3.19 & 2.29 & 5$\times 10^{-3}$ & 2.24 \\

\cline{2-10}

 & $1\ ^1A'' \leftarrow 1\ ^1A'$ & 2.76 & 2.64 & 2.43 & 2.04 & 2.73 & 2.64 & 2.56
& 1$\times 10^{-2}$ & \\

\hline

CH$_2$NO$^-$ & $2\ ^1A' \leftarrow 1\ ^1A'$ & 6.85 & 2.88 & 1.77 & 1.63 & 6.31
& 2.86 & 1.94 & 3$\times 10^{-3}$ & 2.13 \\

\hline

CH$_2$PO$^-$ & $1\ ^1A'' \leftarrow 1\ ^1A'$ & 6.44 & 3.48 & 2.89 & 2.79 & 6.24
& 3.47 & 2.96 & 2$\times 10^{-3}$ & 2.79\\

\cline{2-10}

 & $2\ ^1A' \leftarrow 1\ ^1A'$ & 5.37 & 4.70 & 3.05 & 2.81 & 5.19 & 4.65 & 3.43
& 1$\times 10^{-4}$ & \\

\hline

C$_3$H$^-$
 & $1\ ^1A'' \leftarrow 1\ ^1A'$ & 1.21 & 1.15 & 1.14 & 1.14 & 1.17 & 1.14 &
1.13 & 2$\times 10^{-3}$ & 2.34 \\

\cline{2-10}

 & $2\ ^1A' \leftarrow 1\ ^1A'$ & 6.38 & 3.01 & 2.46 & 2.37 & 5.68 &
3.03 & 2.61 & 3$\times 10^{-3}$ & \\

\hline

CCSiH$^-$
 & $1\ ^1A'' \leftarrow 1\ ^1A'$ & 2.09 & 2.05 & 2.05 & 2.05 & 2.11 & 2.07 &
2.07 & 1$\times 10^{-2}$ & 3.27\\

\cline{2-10}

 & $2\ ^1A' \leftarrow 1\ ^1A'$ & 6.40 &
4.02 & 3.32 & 3.26 & 6.00 & 3.92 & 3.45 & 2$\times 10^{-2}$ & \\

\hline

CCOH$^-$$^d$ & 1$\ ^1A''\leftarrow 1\ ^1A'$ &  4.85 & 2.92 &
2.68 & 2.66 & 4.47 & 2.99 & 2.82 & 3$\times 10^{-3}$ & 2.76 \\

\hline

CCSH$^-$ & $1\ ^1A'' \leftarrow 1\ ^1A'$ & 4.59 & 3.26 & 2.92 & 2.88 & 4.44 &
3.31 & 3.06 & 2$\times 10^{-3}$ & 2.94 \\

\hline

CCNH$_2^-$ & $2\ ^1A' \leftarrow 1\ ^1A'$ & 4.74 & 2.59 & 2.36 & 2.31 & 4.24 &
2.66 & 2.47 & 1$\times 10^{-4}$ & 2.42\\

\hline

CCPH$_2^-$ & $2\ ^1A' \leftarrow 1\ ^1A'$ & 6.82 & 4.59 & 3.72 & 3.27 & 6.11 &
4.56 & 3.44 & 2$\times 10^{-3}$ & 3.30 \\

\hline

BH$_3$NH$_2^-$ & $2\ ^1A' \leftarrow 1\ ^1A'$ & 5.94 & 3.35 & 2.93 & 2.87 & 5.35
& 3.39 & 3.07 & 8$\times 10^{-3}$ & 2.88\\

\hline

BH$_3$PH$_2^-$ & $2\ ^1A' \leftarrow 1\ ^1A'$ & 6.51 & 3.88 & 3.35 & 3.26 &
5.93 & 3.89 & 3.41 & 3$\times 10^{-3}$ & 3.26 \\

\hline

AlH$_3$NH$_2^-$ & $2\ ^1A' \leftarrow 1\ ^1A'$ & --$^e$ & 3.99 & 3.70 & 3.69 &
-- & 4.05 & 3.86 & 1$\times 10^{-2}$ & 3.80 \\

\cline{2-10}

 & $1\ ^1A''^f \leftarrow 1\ ^1A'$ & 7.25 & 4.56 & 3.98 & 3.85 & 6.88 & 4.60
& 4.13 & 1$\times 10^{-5}$ & \\

\hline

AlH$_3$PH$_2^-$ & $2\ ^1A' \leftarrow 1\ ^1A'$ & --$^e$ & 4.22 & 3.84 & 3.80 &
-- & 4.27 & 3.96 & 2$\times 10^{-2}$ & 3.85 \\

\cline{2-10}

 & $1\ ^1A''^f \leftarrow 1\ ^1A'$ & 7.38 & 4.69 & 4.03 & 3.86 & 7.04 & 4.70 &
4.16 & 1$\times 10^{-3}$ & \\ 

\hline
\end{tabular}

$^a$Oscillator strengths ($f$ values) reported are for CCSD/d-aug-cc-pVTZ.
$^b$Dunning's correlation consistent basis sets are abbreviated, {\em e.g.}
t-aug-cc-pVDZ is tapVDZ.
$^c$Computed with EOMIP-CCSD/t-aug-cc-pVDZ.
$^d$Computed by \cite{Fortenberry11dbs}.
$^e$No clear correspondence between the pVDZ/pVTZ results and those with more
diffuse basis sets could be established.
$^f$Part of a degenerate set of 4$p$ Rydberg-like states.

\end{table}

\renewcommand{\baselinestretch}{1}

\end{document}